\documentclass[a4paper]{panl}
\usepackage{seqsplit}
\usepackage{cite}
\usepackage{wrapfig}
\usepackage{graphicx}
\usepackage{amssymb}
\usepackage{amsfonts}
\usepackage{amsmath}
\usepackage{longtable}
\usepackage{rotating}
\usepackage{lscape}
\usepackage{epsfig}
\usepackage{multirow}

\def\bbljan{Jan}
\def\bblfeb{Feb}
\def\bblmar{Mar}
\def\bblapr{Apr}
\def\bblmay{May}
\def\bbljun{Jun}
\def\bbljul{Jul}
\def\bblaug{Aug}
\def\bblsep{Sep}
\def\bbloct{Oct}
\def\bblnov{Nov}
\def\bbldec{Dec}

\def\apj{ApJ}
\def\apjl{ApJL}

\def\jcap{JCAP}
\def\mnras{MNRAS}
\def\nat{Nature}
\def\na{New Astronomy}

\def\physrep{Phys. Rep.}
\def\sovast{Sov. Astron}

\def\prd{Phys. Rev. D}
\def\prl{Phys. Rev. Lett}

\def\M{{\cal M}}
\def\BibDash{--}

\originalTeX
\begin{document}
% Journal sections (see http://pkp.jinr.ru/index.php/PEPAN_LETTERS/about/editorialPolicies#focusAndScope)
%\issuearea{Physics of Elementary Particles and Atomic Nuclei. Theory}
% or in Russian
%\issuearea{ФИЗИКА ЭЛЕМЕНТАРНЫХ ЧАСТИЦ И АТОМНОГО ЯДРА. ТЕОРИЯ}

\title{Gravitational wave astronomy: astrophysical and cosmological inferences}
\maketitle
%\authors{K.A.\,Postnov$^{a,b,}$\footnote{E-mail: first.author@email.ru},S.\,Author$^{b,}$\footnote{E-mail: second.author@email.ru}}
\setcounter{footnote}{0}
\authors{K.A.\,Postnov$^{a,b,}$\footnote{E-mail: pk@sai.msu.ru},
N.A.\,Mitichkin$^{a,}$\footnote{E-mail: mitichkin.nikita99@mail.ru}}
%\from{$^{a}$\,Affiliation 1}
\from{$^{a}$\, Sternberg Astronomical Institute, M.V. Lomonosov Moscow State University, Universitetskij pr. 13, 119234, Moscow, Russia}
\from{$^{b}$\, Department of Physics, Novosibirsk State University, Pirogova 2, 630090, Novosibirsk, Russia}
%\from{$^{b}$\,Место работы автора 2}

%%%%%%%%%%%%% line numbering
%\linenumbers
%%%%%%%%%%%%% line numbering

\begin{abstract}
We briefly discuss the most prominent results and specific sources detected by gravitational-wave observatories LIGO-Virgo during first three O1-O3 runs, as well as possible astrophysical and cosmological channels of their formation. We show that it is possible to explain the observed correlation between the effective spin of coalescing binary black holes and mass ratio of the components by accretion from the ambient medium onto primordial binary black holes. We also briefly discuss the recent results of searches for stochastic gravitational-wave background in the nano-Hz frequency band by pulsar timing arrays.  
\end{abstract}
\vspace*{6pt}

\label{sec:intro}
\section*{Introduction}
Gravitational-wave astronomy is a new and rapidly developing field of science that investigates astrophysical sources of gravitational waves (GW) using methods of multi-wavelength ground-based and space astronomy (from radio to gamma-ray wavelengths). The beginning of GW astronomy started with the registration of the first source, the merging of massive black holes (BHs) in the binary system GW150914 \cite{PhysRevLett.116.061102}. The merger of two BHs as the most likely astrophysical source in the frequency range of sensitivity of the LIGO GW interferometers $\sim 10-1000$~Hz was anticipated from the evolution of binary stars \cite{1993MNRAS.260..675T, 1997MNRAS.288..245L, 1997NewA....2...43L, 1997AstL...23..492L,2005PhyU...48.1235G}. The first detection of GW from a merging system of two neutron stars (NSs) GW170817 \cite{PhysRevLett.119.161101} was associated with a short gamma-ray burst GRB170817A (anticipated by the physics of NS+NS mergings, e.g. \cite{1984SvAL...10..177B,1989Natur.340..126E})
and accompanied by a follow-up multi-wavelength electromagnetic signal  \cite{2017ApJ...848L..12A} arising from thermal and non-thermal emission of matter during the merger (the so-called "kilonova" \cite{2010MNRAS.406.2650M}). The main results of the LVK (LIGO-Virgo-Kagra) collaboration for the past three O1-O3 observing seasons are presented in the GWTC-3 \cite{2021arXiv211103606T} catalogue, including $\sim 90$ sources of astrophysical origin -- candidates for merging binary BH, BH+NS and NS+NS. Hundreds of papers have already been devoted to the analysis of these data and their interpretation (see, for example, reviews \cite{2022Galax..10...76S,2022LRR....25....1M} and references therein). No GW-signals other than from merging double compact systems have been reliably detected so far (see, however, the \ref{s:PTA} section below).

\section{Binary BHs detected in GW observations}
Analysis of GW observations of merging binary systems with masses $m_1$ and $m_2$ and dimensionless moments (spins) $a_i^*=J_ic/(Gm_i^2)$ ($G$, $c$ -- Newtonian gravitational constant and light speed) enables estimating (with varying accuracy depending on the signal amplitude) the following parameters: (1) the chirp-mass $\M=\frac{(m_1m_2)^{3/5}}{(m_1+m_2)^{1/5}}$ in the observer system $\M_\mathrm{det}$\footnote[1]{Chirp-mass in the observer system is determined by measuring signal frequency during the inspiraling along quasi-Keplerian orbits before merging and is degenerate by source redshift $z$ $\M_\mathrm{det}=(1+z)\M$}, the masses of the individual components and the mass ratio $q=m_2/m_1\le 1$, (2) the effective spin before merging $\chi_{\mathrm{eff}}=\frac{m_1a_1^*\cos\theta_1+m_2a_2^*\cos\theta_2}{m_1+m_1}$ (mass-weighted projection of individual spins on the orbital momentum), the projection of the total spin of the components on the orbital plane $\chi_p$, the spin of the resulting object after the merger $a^*_\mathrm{fin}$, (3) the photometric distance to the source $D_l(z)$ (redshift $z$ in a given cosmological model) \cite{1986Natur.323..310S}. The classification of the merging binary system type by the GW signal is usually based on the estimation of the component masses -- a compact object with mass $m>3 M_\odot$ is known to be a BH \cite{1974PhRvL..32..324R,2016PhR...621..127L}{\footnote[2]{Maximal known mass of a NS in millisecond pulsar PSR J0952-0607 is $M_{PSR}\approx 2.35 M_\odot$ \cite{2022ApJ...934L..17R}.}. The source type  can be also identified from the GW signal alone by machine learning methods \cite{Qiu:2022wub}.

An analysis of the observed parameters of merging binary BHs detected at GW interferometers by the LVK Collaboration leads to several important astrophysical conclusions.
\begin{enumerate}
    \item     
    There exist BHs with tens of solar masses which are different in their characteristics (masses, spins) from BHs observed in Galactic X-ray close binary systems. The evolution of single massive stars predicts a mass gap in BH masses $\sim 60 M_\odot-120 M_\odot$ caused by pulsational pair instability due to electron-positron pair production in the stellar core, leading to a pair-instability supernovae explosion \cite{2017ApJ...836..244W,2021ApJ...912L..31W}.  However, the source GW190521 was found to have BH masses before merging in the mass gap: $m_1\approx 85 M_\odot, m_2\approx 66 M_\odot$ \cite{PhysRevLett.125.101102}. It is possible to obtain such BH masses in the evolution scenario of isolated massive binary systems only using special assumptions \cite{2020ApJ...902L..36F}. A possible channel for the formation of binary BHs with large masses may be the dynamic evolution in dense stellar clusters with hierarchical growth of BH mass and spin in successive mergers \cite{2021ApJ...915L..35K}. Also, the formation of such BH+BH from population III stars with primordial chemical composition \cite{2021ApJ...910...30T} and in the vicinity of active galaxy nuclei \cite{2021ApJ...908..194T} is discussed.
    
    \item
    Large (almost limiting) spins of accreting BHs in Galactic X-ray binary systems \cite{2022arXiv221002479D} are measured. Such spinning-up of BHs is natural via accretion of matter from the second component. In binary BHs formed via the standard astrophysical channel from the evolution of massive binary systems, large effective spins are possible under certain additional assumptions \cite{2019MNRAS.483.3288P}. In the source GW200129, the shape of the inspiraling GW signal evidences for an orbital relativistic precession, indicating a record high spin of the primary (more massive component with mass $m_1\approx 39 M_\odot$) $a_1/m_1=0.9^{+0.1}_{-0.5}$, inclined to the orbital angular momentum \cite{Hannam:2022pit}. It is difficult to produce such sources in the most popular astrophysical scenarios of binary BH formation from isolated massive binary systems \cite{2016Natur.534..512B} or via dynamic captures in dense star clusters \cite{2016ApJ...824L...8R}. The mechanism of formation of similar binary BHs in the vicinity of active galactic nuclei is discussed in \cite{2021MNRAS.507.3362T}.

%============================= Fig. 1 ================================
\begin{figure}[t]
\begin{center}
\includegraphics[width=127mm]{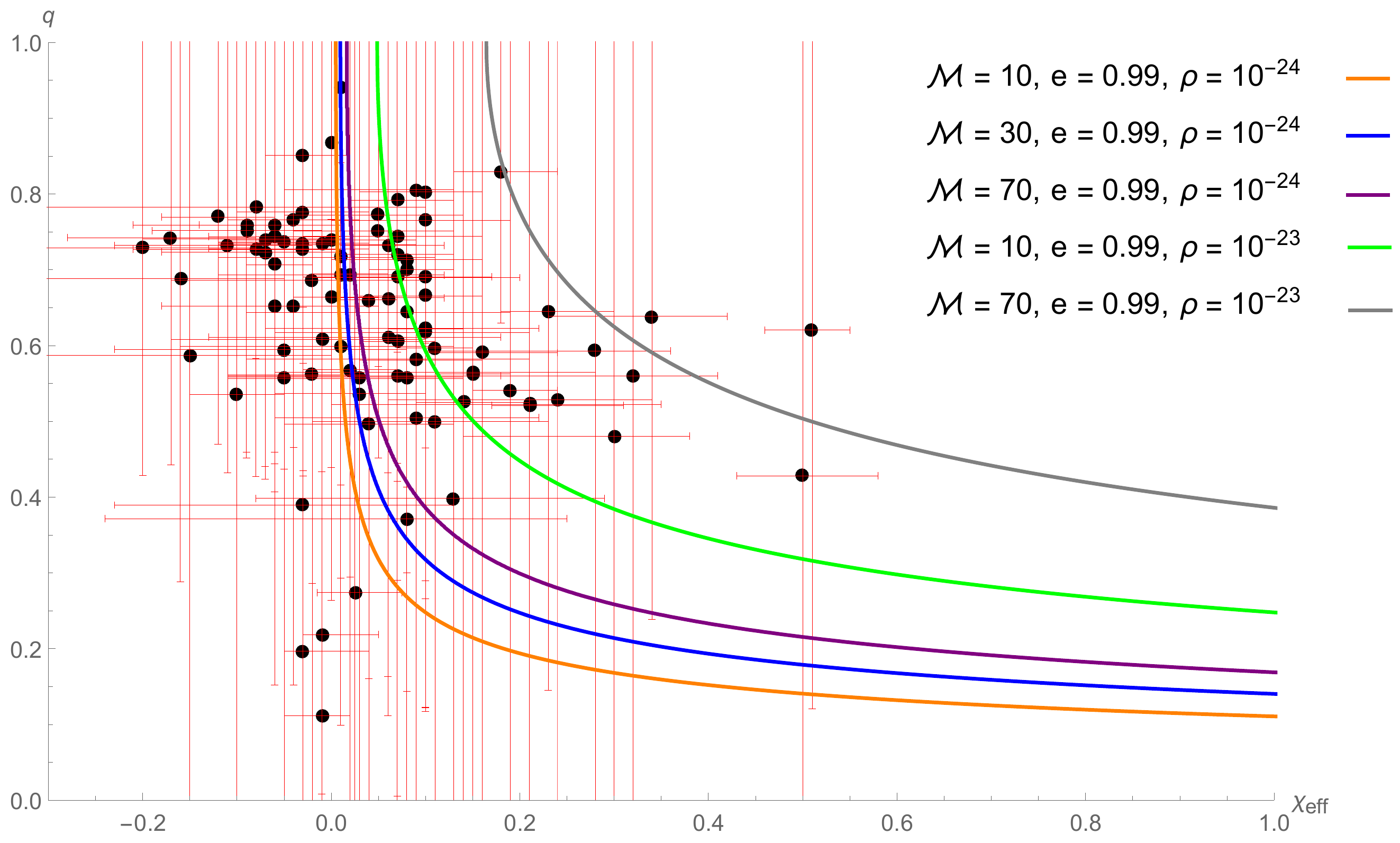}
\vspace{-3mm}
\caption{The mass ratio $q=m_2/m_1$ and effective momentum $\chi_{eff}$ of merging binary BHs from the GWTC-3 \cite{2021arXiv211103606T} catalogue. The colored curves are the expected $\chi_{eff}-q$ correlations due to accretion from the interstellar matter onto a PBH with component mass ratio $q$ in the model \cite{2019JCAP...06..044P}.}
\end{center}
\label{f:1}
\vspace{-5mm}
\end{figure}
%============================= Fig. 1 ================================
    
    \item
    Possible indications of primary stellar mass BHs (PBHs) from GW observations \cite{2016PhRvL.116t1301B,2016JCAP...11..036B,2019JCAP...05..018D}. In the standard mechanism of PBH formation \cite{1967SvA....10..602Z,1974MNRAS.168..399C}, the masses of the PBHs can naturally be of the order of 10 $M_\odot$ (of the order of mass inside the cosmological horizon at the QCD phase transition epoch at $T\sim 150$ MeV) \cite{2020JCAP...07..063D} with a log-normal mass spectrum predicted in some models \cite{1993PhRvD..47.4244D}. The log-normal mass spectrum of PBHs is most likely in explaining (at least a part of) the observed LIGO-Virgo sources \cite{2020JCAP...12..017D,Liu:2022iuf}. Indications of a significant correlation of the effective spin of merging BH+BH with the mass ratio $q$ were found \cite{2021ApJ...915L..35K,2022MNRAS.tmp.2764A}. In the simplest models, the initial spin of PBHs should be close to zero, but in binary PBHs it can increase due to accretion of the surrounding baryonic matter \cite{2019JCAP...06..044P}. Fig. \ref{f:1} shows the effective spins $\chi_{eff}$ and mass ratios $q=m_2/m_1$ of 84 double BHs from the GWTC-3 \cite{2021arXiv211103606T} catalogue. The curves demonstrate the expected correlations of the effective spin of merging binary PBHs for different values of chirp masses ${\cal M}$, orbital eccentricities at the formation and ambient matter density \cite{2019JCAP...06..044P}. It can be seen that in dense environments (e.g. gas disks in galactic nuclei) the PBHs can acquire high effective spins (upper two curves). The question about the fraction of PBHs among the observed binary BH+BH sources detected by the LVK collaboration remains open and is actively discussed in the literature (see, for example, \cite{2022arXiv220906196E,Liu:2022iuf}). 
\end{enumerate}

\section{Binary NS+NS and NS+BH in GW observations}
The GW170817 binary NS merger, accompanied by multi-wavelength EM radiation (short gamma-ray burst GRB170817A and kilonova AT 2017gfo in NGC4993 galaxy), has provided the richest source information and a number of fundamental physical and cosmological constraints \cite{2017ApJ...848L..12A,2017ApJ...850L..24G}. In addition to the masses and spin of the components, the analysis of the GW signal from NS+NS imposes constraints on the tidal strain parameters, which in turn depend on the NS equation of state. Constraints on the equation of state and the upper NS mass (Tolman-Oppenheimer-Volkov limit, $M_{TOV}$) from observations of merging double NSs GW170817, GW190814 and (possible candidate) GW190425 see, e.g., \cite{2018ApJ...852L..25R,2022ApJ...926...75B}.

The electromagnetic signal from merging binary NS+BH is expected in the case of a small mass ratio of the components (i.e., for sufficiently light BHs) \cite{2020AstL...45..728P}. In BH+NS candidates, the masses of the components and the spin of the BH are only estimated from GW observations. Based on the small mass of the second component $<M_{TOV}$, sources GW200105 and GW200115 are classified as
BH+NS mergers, \cite{2021ApJ...915L...5A}. The estimate of the merger rate and the fraction of registered NS+BH does not contradict expectations from the analysis of the evolution of massive binary systems \cite{2019PhyU...62.1153P}.

\section{Searching for stochastic GW background using the pulsar timing}
\label{s:PTA}
The pulsar timing method of GW detection was proposed in the late 1970s \cite{1978SvA....22...36S,1979ApJ...234.1100D} and is used to search for a possible stochastic GW background from analysis of long-year observations by several ground-based radio telescope collaborations (pulsar timing arrays) in the nano-Hz frequency range (see review \cite{2021hgwa.bookE...4V} and references therein). In the pulsar timing method, a GW with amplitude $h$ at frequency $f$ causes residual deviations in pulse arrival times $r\simeq 10[ns] (h/10^{-16})/(f/10^{-8}\mathrm{Hz})$, which is close to the current sensitivity limit of arrival time measurements from millisecond pulsars. The characteristic amplitude of the stochastic GW background $h_c$ is usually sought as a power law $h_c=A(f/1 \mathrm{yr}^{-1})$. A stochastic GW background can arise, in particular, during the merger of supermassive binary BHs in the nuclei of galaxies \cite{2008MNRAS.390..192S} ($\alpha=-2/3$), can be generated by cosmic strings \cite{2010PhRvD..81j4028O} ($\alpha=-7/6$), during phase transitions in the early Universe \cite{2005PhyU...48.1235G} ($\alpha=-1$) from primary cosmological perturbations in the inflationary stage \cite{1975JETP...40..409G,1982PhLB..115..189R} ($\alpha=-1$). A GW background can also accompany the formation of PBHs  collapsing from primordial scalar perturbations \cite{2021PhRvL.126d1303D}. Recently, 
NANOGrav \cite{2020ApJ...905L..34A} and 
IPTA \cite{2022MNRAS.510.4873A} collaborations reported the presence of a common signal from a stochastic process with amplitude $A\approx 4\times 10^{-15}$ and slope $\alpha \sim -0.5$ in the data, which can be interpreted, in particular, as a GW background from supermassive BHs with a spatial density of $\sim 10^{-5}$ Mpc$^{-3}$ \cite{2022MNRAS.510.4873A}. However, to prove the GW nature of this signal, it is necessary to find a specific spatial quadrupole Hellings-Downes correlation \cite{1983ApJ...265L..39H}, which has not yet been reliably established. Confirmation of the discovery of a stochastic GW background  
in the pulsar timing data will be another milestone experimental result of GW astronomy.

\vskip\baselineskip
\textit{Acknowledgments.} The authors acknowledge the support from the Russian Science Foundation grant 22-12-00103.

 %\cite{Silenko:2014yxa,Bogolyubov:1983gp,hanson-67,wright-63,Anselmino:2008jk,Aad:2012tfa,Beda:2009kx,Franceschini:2015kwy} are as follows

%\nocite{*}
%\bibliographystyle{pepan}
%\bibliography{GW1}

\end{document}